\documentclass[letterpaper, 10 pt, conference]{ieeeconf}  

\IEEEoverridecommandlockouts                              

                                                          \usepackage{blindtext}
\usepackage{graphicx}
\usepackage[utf8]{inputenc}
\usepackage{graphics} 
\usepackage{epsfig} 
\usepackage{mathptmx} 
\usepackage{amsmath} 
\usepackage{amssymb}  
\usepackage{cite}
\usepackage{physics}
\usepackage{multicol}
\usepackage{mathtools, cuted}
\usepackage[cmintegrals]{newtxmath}
\usepackage{epstopdf}
\usepackage{graphics} 
\usepackage{epsfig} 

\usepackage{mathptmx} 
\usepackage{amsmath} 
\usepackage{amssymb}  
\usepackage{cite}
\usepackage{multicol}
\usepackage{mathtools, cuted}
\usepackage[cmintegrals]{newtxmath}
\usepackage{mathtools}

\usepackage{mathptmx}
\usepackage{helvet}
\usepackage{courier}
\usepackage{subfigure}
\usepackage{type1cm}
\usepackage{titlesec}
\usepackage{epsfig} 
\usepackage{mathptmx} 
\usepackage{amsmath} 
\usepackage{amssymb}  
\usepackage{cite}
\usepackage{multicol}
\usepackage{mathtools, cuted}
\usepackage[cmintegrals]{newtxmath}
\usepackage{makeidx}         
\usepackage{algorithm}
\usepackage{algpseudocode}

\usepackage{multicol}        
\usepackage[bottom]{footmisc}
\usepackage{caption}
\usepackage{nomencl}
\usepackage{soul}
\usepackage[usenames, dvipsnames]{color}
\usepackage{url}
\usepackage{siunitx}

\usepackage{multicol}        
\usepackage[bottom]{footmisc}
\usepackage{caption}
\usepackage{titlesec}
\usepackage{nomencl}
\usepackage[usenames, dvipsnames]{color}
\usepackage{multirow}
\usepackage{adjustbox}
\usepackage{amsthm}
 
\theoremstyle{definition}
\newtheorem{definition}{Definition}

\theoremstyle{remark}
\newtheorem{remark}{Remark}

\title{\LARGE \bf
Online Equitable Ride-sharing Car Distribution in a Congested Traffic
}

\author{Hossein Rastgoftar
\thanks{Author is with the Aerospace and Mechanical Engineering Department,
 University of Arizona, Tucson,
AZ, 85719 USA e-mail: \{hrastgoftar\}@arizona.edu.}
}

\begin{document}

\maketitle

\begin{abstract}
The usability of ride-sharing services like Uber and Lyft has been considerably improved by advancements in cellular communications. Such a tech-driven transportation system can reduce the number of private cars, in roads with limited physical capacity, effectively match drivers with passengers requesting service, and save drivers' and passengers' time by optimal scheduling and estimating of the trips. However, the existing services offered by ride-sharing businesses may not necessarily reduce congestion, if they are not equitably accessed in urban areas. This paper addresses this important issue by classifying cars as ride-sharing and non-ride-sharing cars and developing a novel dissensus-based traffic evolution dynamics to effectively model their distribution in a shared network of interconnected roads (NOIR). This new dynamics models the ride-sharing car evolution as a non-stationary Markov process with uncertainty matrices decomposed into outflow and tendency probability matrices, where outflow probabilities are constrained by the non-ride-sharing cars' behavior, but the tendency probabilities are considered as decision variables (or distributed controls) for achieving an equitable distribution of ride-sharing cars in a congested traffic.


\end{abstract}

\section{Introduction}
Ride-sharing services like Uber and Lyft are extremely popular these days. Passengers can economically benefit from using ride-sharing cars. Passengers using ride-sharing services save time and avoid potential distraction for finding destination addresses, and they do not need to search for parking in congested urban areas. Utilizing ride-sharing services can also reduce the total number of private vehicles in urban areas, which will help to mitigate the effects of traffic congestion. Despite these advantages, not everyone who wants to travel in cities equally access ride-sharing services.  The main contribution of this paper is to provide a framework for online equitable distribution of ride-sharing cars in a congested traffic.

\subsection{Related Work}
Road traffic congestion modelling and control has been an active research area over the past few decades. Light-based approaches use model-free strategies to obtain an optimised timing strategy for traffic signals, whereas physics-based approaches model the traffic system by the utilisation of queue theory and traffic flow, as Ekeocha and VI Ihebom demonstrated in their work\cite{queue_trafflow}. Moreover, to divide a NOIR into road elements, a notable number of publications implemented the Cell Transmission Model (CTM) to obtain a traffic coordination modeling \cite{CTM1, CTM2, CTM3}. 

On the other hand, the traffic congestion could be modelled as a MIMO\footnote{Multiple-input, multiple-output} boundary control system, so that the model could be treated as a quadratic convex optimisation problem with multiple constraints that maximise the total vehicle flow efficiency \cite{IR_BoundaryControl}, while ensuring the feasibility of traffic coordination by the implementation of model-predictive control (MPC)\cite{drrastgoftar(ANZCC)}.

Aligned with the growing applications of artificial intelligence (AI) and data-driven approaches \cite{datadriven_opt} over the past decades, a significant number of proposals have been allocated to implement these algorithms to solve the corresponding optimisation problem, i.e., Support Vector Machine \cite{SVM} and LQV\cite{LQVNN} neural networks, deep convolutional neural networks \cite{Deep}, etc. Additionally, the traffic congestion could be considered as a consensus model integrated with MPC strategy \cite{MPCconsensus},  Fuzzy Evidential Reasoning Algorithm (FERA) \cite{FERAconsensus}, etc., where the system states are expected to finally converge to fixed desired states.

Recently, some works have been focusing on the coverage of ride-sourcing systems to contribute to the high-demanding for efficient urban transportation  \cite{coverage172}, as well as proposing various optimisation pricing strategies, i.e., electric vehicle fuel charging management \cite{chargepricing1, chargepricing2}. Associated with the recent works, traffic congestion management could be categorised into two major aspects: physics-based analysis \cite{rastgoftar2019physics} and light-based analysis \cite{lightbased}.

\subsection{Contributions}
We develop a weighted dissensus dynamics to properly integrate ride-sharing cars into traffic coordination modeling. In contrast to the existing consensus model \cite{ren2005survey}, which guarantees that a network system's state components all converge to a fixed desired state, our dissensus-based evolution model allows a network to achieve a desired distribution of states.  We will use this aspect of the  dissensus coordination to achieve an equitable distribution of ride-sharing cars in a congested traffic.  

The proposed dissensus-based dynamics models traffic evolution as a non-stationary Markov process. This enables to quantify uncertain weights of the Markov process based on ``outflow'' and ``tendency'' probabilities \cite{rastgoftar2022concurrent}, defined for ride-sharing   and \textit{all} cars, by applying the  cell transmission model  (CTM). More specifically,  we consider the  ride-sharing tendency probabilities as decision variables that can be commanded by ride-sharing  cars subject to inequality and equality feasibility constraints obtained by imposing the CTM.  We develop a quadratic programming to plan the  ride-sharing cars tendency probabilities in a congested traffic, assuring an equitable distribution in a congested traffic

\subsection{Outline}
This paper is organized as follows: Section \ref{Preliminary Notions of Graph Theory} presents preliminary notions on graph theory and dissensus-based network evolution. The problem of equitable ride-sharing car distribution is stated and formulated in Section \ref{Problem Statement}. Equitable traffic is modeled as  discensus coordination in Section \ref{Dissensus-Based Traffic Evolution}. We formulate a quadratic programming in Section \ref{Equitable Traffic Management} to achieve equitable distribution of ride-sharing in a congested traffic, in real-time. Simulation results are presented in Section \ref{Simulation Results} and followed by Conclusion in Section \ref{Conclusion}.



\section{Preliminary Notions}\label{Preliminary Notions of Graph Theory}
\subsection{Graph Theory Notions}
We propose to represent a network of interconnected roads (NOIR) by graph $\mathcal{G}\left(\mathcal{V},\mathcal{E}\right)$, where  node set $\mathcal{V}=\{1,\cdots,N\}$ defines the identification numbers of the roads, and edge set $\mathcal{E}\subset \mathcal{V}\times \mathcal{V}$ specifies interconnections between the NOIR roads. If $(i,j)\in \mathcal{E}$, we say traffic is flowing from $i\in \mathcal{V}$ to $j\in \mathcal{V}$. We can specify in-neighbor road set $\mathcal{I}_{i}=\{j|\left(j,i\right)\subset\mathcal{E}\}$ and the out-neighbor road set $\mathcal{O}_{i}=\{j|\left(i,j\right)\subset\mathcal{E}\}$, for every road element $i\in \mathcal{V}$, given edge set $\mathcal{E}$. In particular, traffic is flowing from $j\in \mathcal{I}_i$ to $i\in \mathcal{V}$. We can also say that traffic is flowing for $i\in \mathcal{V}$ to $j\in \mathcal{O}_i$. 


\subsection{Dissecus-Based Communication (Laplacian) Matrix}
Given  graph $\mathcal{G}\left(\mathcal{V},\mathcal{E}\right)$, we can define time-varying discensus-based  communication matrix $\mathbf{A}_k=\left[A_k^{i,j}\right]\in \mathbb{R}^{N\times N}$ with $(i,j)$ entry
\begin{equation}
    A_k^{i,j}=\begin{cases}
        a_k^{i,j}\geq 0&\left(\left(i,j\right)\in \mathcal{E}\vee j=i \right)\wedge \left(i\in \mathcal{V}\right)\\
        0&\left(\left(i,j\right)\notin \mathcal{E}\wedge j\neq i\right)\wedge \left(i\in \mathcal{V}\right)\\
    \end{cases}
    ,\qquad \forall k\in \mathbb{N},
\end{equation}
where $\wedge$ and $\vee$ denote logical ``and'' and logical ``or'', respectively, and $k$ denotes discrete time. Communication weight $a_k^{i,j}$ satisfies  
\begin{equation}
\sum_{i=1}^Na_k^{i,j}=1,\qquad \forall k\in \mathbb{N},~\forall j\in \mathcal{V},
\end{equation}
which in turn implies that columns of matrix $\mathbf{A}_k$ all sum up to $1$.


\section{Problem Statement}\label{Problem Statement}
We consider an NOIR represented by graph $\mathcal{G}\left(\mathcal{V},\mathcal{E}\right)$ and apply CTM to model traffic evolution. Therefore, we define 
 $\mathbf{x}_k=\begin{bmatrix}
    x_k^1&\cdots&x_k^N
\end{bmatrix}^T$ as the state vector aggregating traffic density of \textit{all} cars at every NOIR road, and model traffic dynamics by 
\begin{equation}\label{allcardynamics}
\mathbf{x}_{k+1}=\mathbf{A}_k\mathbf{x}_{k}+\mathbf{d}_k,
\end{equation}
where $\mathbf{A}_k\in \mathbb{R}^{N\times N}$ is a non-negative dissensus-based communication matrix and $\mathbf{d}_k=\begin{bmatrix}
    d_{k}^1&\cdots&d_{k}^{N}
\end{bmatrix}$ aggregate the exogenous cars that either enter or leave the NOIR. 

Similarly, we apply CTM to model evolution of ride-sharing cars. 
By defining $\hat{x}_k^i$ as number of existing free ride-sharing (\textit{FRS}) cars in road $i\in \mathcal{V}$, that are not matched with any passenger at discrete time $k$, $\hat{\mathbf{x}}_k=\begin{bmatrix}
    x_k^1&\cdots&x_k^N
\end{bmatrix}^T$ is updated by 
\begin{equation}\label{frscarsdynamics}
\hat{\mathbf{x}}_{k+1}=\hat{\mathbf{A}}_k\mathbf{x}_{k},
\end{equation}
where $\hat{\mathbf{A}}_k$ is a non-negative matrix. 
\begin{remark}
    Unlike the consensus evolution at which sum of the rows of the network matrix is $1$,  in our proposed dissensus evolution, sum of every  column of matrices $\mathbf{A}_k$ and $\hat{\mathbf{A}}_k$ is $1$.
\end{remark}
Because \textit{FRS} cars are  a portion \textit{all} cars at every road $i\in \mathcal{V}$, $\hat{\mathbf{x}}_k$  and $\mathbf{x}_k$ must satisfy the following inequality constraint:
   \begin{equation}\label{inequalityupper}
       \hat{\mathbf{x}}_k\leq {\mathbf{x}}_k,\qquad \forall k,
    \end{equation}
Furthermore, we desire that there exists at least $\hat{x}_{i,min}$ \textit{FRS} cars at every road $i\in \mathcal{V}$ to ensure their equitable distribution in a congested traffic. This requirement is imposed by defining inequality constraint 
    \begin{equation}\label{inequality}
         \hat{\mathbf{x}}_k\geq \hat{\mathbf{x}}_{min},\qquad \forall k,
    \end{equation}
where the lower-bound  $\hat{\mathbf{x}}_{min}=\begin{bmatrix}\hat{x}_{1,min}&\cdots&\hat{x}_{N,min}\end{bmatrix}^T\geq \mathbf{0}$ is known. 

Given above traffic evolution modeling, this paper studies the following two main problems:

\textbf{Problem 1:} We model traffic evolution as non-stationary Markow process that is specified based on the outflow and tendency probabilities for  \textit{\textit{FRS}} and \textit{all} cars using the NOIR. Section \ref{Dissensus-Based Traffic Evolution} applies CTM to decompose matrix $\mathbf{A}_k$, based on outflow and tendency probabilities of \textit{all} cars,  and $\hat{\mathbf{A}}_k$, based on outflow and tendency probabilities of \textit{FRS} cars.

\textbf{Problem 2:} We consider the non-zero entries of matrix $\hat{\mathbf{A}}_k$ as decision variables that can be partially controlled by the ride-sharing cars but partially constrained by the inequality constraint \eqref{inequality}. We formulate this problem as a quadratic problem in Section \ref{Equitable Traffic Management}. Given updated  $\hat{\mathbf{A}}_k$, at every discrete time $k$, Section \ref{Equitable Traffic Management} also formulates a quadratic programming problem for updating $\mathbf{A}_k$ at every discrete time $k$.

\section{Dissensus-Based Traffic Evolution}\label{Dissensus-Based Traffic Evolution}
In this section, we explain how dissensus matrices $\mathbf{A}_k$ and $\hat{\mathbf{A}}_k$ can be quantified at every discrete time $k$ by applying the CTM. To this end, for every road $i\in \mathcal{V}$ and discrete time $k$, we use:
\begin{enumerate}
\item $\hat{x}_k^i$, $\hat{y}_k^i$, and $\hat{z}_k^i$  denote the traffic density, the network inflow, and the network outflow of the \textit{FRS} cars.
    \item $x_k^i$, $y_k^i$, and $z_k^i$  denote the traffic density, the network inflow, and the network outflow of \textit{all} cars.     
\end{enumerate}


By applying the available CTM, the dynamics of \textit{all} and \textit{FRS} cars are given by
\begin{subequations}\label{rawdynamics}
    \begin{equation}\label{rawdynamicsa}
    x_{k+1}^i=x_{k}^i+y_{k}^i-z_{k}^i+d_{k}^i,\qquad \forall i\in \mathcal{V},~k\in \mathbb{N},
\end{equation}
    \begin{equation}\label{rawdynamicsb}
    \hat{x}_{k+1}^i=\hat{x}_{k}^i+\hat{y}_{k}^i-\hat{z}_{k}^i,\qquad \forall i\in \mathcal{V},~k\in \mathbb{N},
\end{equation}
\end{subequations}
where $d_k^i$ is the total number of exogenous cars entering or leaving the road $i\in \mathcal{V}$. By adopting the model proposed in the author's previous work \cite{rastgoftar2021physics, rastgoftar2022concurrent}, we define 
 \begin{subequations}
    \begin{equation}\label{outflowall}
        z_k^i=p_k^ix_k^i,\qquad i\in \mathcal{V},~k\in \mathbb{N}.
    \end{equation}
    \begin{equation}
        \hat{z}_k^i={p}_k^i\hat{x}_k^i,\qquad i\in \mathcal{V},~k\in \mathbb{N}.
    \end{equation}
\end{subequations}
where $p_k^i\in \left(0,1\right]$, defining the fraction of cars leaving road $i\in \mathcal{V}$ at time $k$, is called \textit{outflow probability}. 
Note that $p_k^i$ is considered the same for both {\textit{FRS}} and \textit{all} cars. This is a legitimate assumption since \textit{FRS} and \textit{all} cars follow the traffic speed at every discrete time $k$.

The network inflows, for \textit{all} and \textit{\textit{FRS}} cars,  are defined  by
\begin{subequations}
    \begin{equation}
        y_k^i=\sum_{j\in \mathcal{I}_i}q_k^{i,j}z_k^j,\qquad i\in \mathcal{V},~k\in \mathbb{N},
\end{equation}
 \begin{equation}
        \hat{y}_k^i=\sum_{j\in \mathcal{I}_i}\hat{q}_k^{i,j}\hat{z}_k^j,\qquad i\in \mathcal{V},~k\in \mathbb{N},
\end{equation}
\end{subequations}
where $q_k^{i,j}$ and $\hat{q}_k^{i,j}$ the fractions of $z_k^i$ and $\hat{z}_k^i$, respectively, that moves toward  road $i\in\mathcal{O}_j$ from $j\in \mathcal{V}$. Therefore,
\begin{subequations}
    \begin{equation}
        \sum_{i\in \mathcal{O}_j}q_k^{i,j}=1,\qquad j\in \mathcal{V},~k\in \mathbb{N},
    \end{equation}
     \begin{equation}
        \sum_{i\in \mathcal{O}_j}q_k^{i,j}=1,\qquad j\in \mathcal{V},~k\in \mathbb{N}.
    \end{equation}
\end{subequations}
We call $q_k^{i,j}$ and $\hat{q}_k^{i,j}$ tendency probabilities for \textit{all} and \textit{FRS} cars, respectively. 

\begin{definition}
    We define \textit{all} tendency probability matrix $\mathbf{Q}_k$ with $(i,j)$ entry
    \begin{equation}
        Q_k^{i,j}=\begin{cases}
            q_k^{i,j}&j\in \mathcal{V}\wedge i\in \mathcal{O}_j\\
            0&\mathrm{otherwise}\\
        \end{cases}
        ,\qquad k\in \mathbb{N}.
        .
    \end{equation}
\end{definition}
\begin{definition}
    We define \textit{FRS} tendency probability matrix $\hat{\mathbf{Q}}_k$ with $(i,j)$ entry
    \begin{equation}
        \hat{Q}_k^{i,j}=\begin{cases}
            \hat{q}_k^{i,j}&j\in \mathcal{V}\wedge i\in \mathcal{O}_j\\
            0&\mathrm{otherwise}\\
        \end{cases}
        ,\qquad k\in \mathbb{N}.
    \end{equation}
\end{definition}
\begin{definition}
    We define outflow probability matrix $\hat{\mathbf{P}}_k$ with $(i,j)$ entry
    \begin{equation}
        \mathbf{P}_k=\mathbf{diag}\left(p_k^1,\cdots,p_k^N\right)\in \mathbb{R}^{N\times N},\qquad k\in \mathbb{N}.
    \end{equation}
\end{definition}
Note that matrix $\mathbf{P}_k$ is diagonal and positive definite at every discrete time $k$.

Given matrics $\mathbf{P}_k$, $\mathbf{Q}_k$, and $\hat{\mathbf{Q}}_k$, at every discrete time $k$, matrices $\mathbf{A}_k$ and $\hat{\mathbf{A}}_k$ are obtained by
\begin{subequations}
    \begin{equation}\label{Ak}
        \mathbf{A}_k=\mathbf{I}_N+\left(\mathbf{Q}_k-\mathbf{I}_N\right)\mathbf{P}_k\in \mathbb{R}^{N\times N},
    \end{equation}
    \begin{equation}\label{ahk}
        \hat{\mathbf{A}}_k=\mathbf{I}_N+\left(\hat{\mathbf{Q}}_k-\mathbf{I}_N\right){\mathbf{P}}_k\in \mathbb{R}^{N\times N},
    \end{equation}
\end{subequations}
if we use traffic the CTM, given by Eqs. \eqref{rawdynamicsa} and \eqref{rawdynamicsb}, to model traffic evolution, where $\mathbf{I}_N\in \mathbb{R}^{N\times N}$ is the identity matrix.

\textbf{Feasibility Constraints on FRS Car Density Distribution:} Because constraint  \eqref{inequalityupper} is satisfied at every discrete time $k$, we conclude that $\hat{\mathbf{x}}_{k+1}\leq \mathbf{x}_{k+1}$ holds at discrete time $k+1$. By substituting  Eqs. \eqref{Ak} and \eqref{ahk} into Eqs. \eqref{allcardynamics} and \eqref{frscarsdynamics}, the constraint equation \eqref{inequalityupper} can be rewritten as follows:
\begin{equation}\label{detailedinequality1}
    \hat{\mathbf{Q}}\mathbf{P}\hat{\mathbf{x}}_k\leq \left(\mathbf{I}_N-\mathbf{P}_k\right)\left(\mathbf{x}_k-\hat{\mathbf{x}}_k\right)+\mathbf{Q}_k\mathbf{P}_k\mathbf{x}_k+\mathbf{d}_k.
\end{equation}

Furthermore, constraint  \eqref{inequality} implies that $\hat{\mathbf{x}}_{k+1}\geq \hat{\mathbf{x}}_{min}$ holds at  discrete time $k+1$, as well. By substituting Eq.  \eqref{ahk} into dynamics \eqref{frscarsdynamics}, constraint equation \eqref{inequality} requires the satsifaction of the following inequality at every discrete time $k$:
\begin{equation}\label{detailedinequality2}
    -\hat{\mathbf{Q}}_k{\mathbf{P}}_k\hat{\mathbf{x}}_k\leq \left(\mathbf{I}_N-\mathbf{P}_k\right)\hat{\mathbf{x}}_k-\hat{\mathbf{x}}_{min}.
\end{equation}

\section{Online Equitable ride-sharing Car Distribution}\label{Equitable Traffic Management}
To achieve an equitable distribution of FRS cars, we consider $\hat{q}_k^{i,j}$ as decision variables and formulate a quadratic problem to obtain them at every discrete time $k$.  To this end, we first provide  the constraints in Section \ref{Constraints} and then formulate the optimization problem in Section \ref{Optimization}.

\subsection{Constraints}\label{Constraints}
To ensure feasibility of our model, $\hat{q}_k^{i,j}$ must satisfy   linear equality and inequality constraints that are specified below.


\subsubsection{Constraint on \textit{FRS} Traffic Density} 
By imposing constraint equation \eqref{detailedinequality1}, we obtain the following inequality constraints on $\hat{q}_k^{i,j}$:
\begin{equation}\label{upperconst}
   \bigwedge_{i\in \mathcal{V}}\bigwedge_{j\in \mathcal{I}_i} \left(e_k^{j}\hat{q}_k^{i,j}\leq h_k^{i}\right),\qquad k\in \mathbb{N},
\end{equation}
where 
\begin{subequations}
\begin{equation}
        e_k^{j}=p_k^{j}\hat{x}_k^{j},
    \end{equation}
\begin{equation}
    h_k^{i,j}=\left(1-p_k^i\right)\left(x_k^i-\hat{x}_k^i\right)+q_k^{i,j}p_k^jx_{k}^j+d_k^i.
\end{equation}
\end{subequations}
Furthermore, constraint equation \eqref{detailedinequality2} requires $\hat{q}_k^{i,j}$ to satisfy the following inequality constraint:
\begin{equation}\label{lowerconst}
   \bigwedge_{i\in \mathcal{V}}\bigwedge_{j\in \mathcal{I}_i} \left(-e_k^{i,j}\hat{q}_k^{i,j}\leq s_k^{i}\right),\qquad k\in \mathbb{N},
\end{equation}
where 
 \begin{equation}
    s_k^{i}=\left(1-p_k^{i}\right)\hat{x}_k^{i}-\hat{x}_{i,min}.
\end{equation}

\subsubsection{Feasibility Constraints on Tendency Probabilities} Tendency probabilities of \textit{\textit{FRS}} cars are always non-negative. Therefore,   $\hat{q}_k^{i,j}$ satisfies the following inequality constraints:
\begin{equation}\label{postiveqh}
    \bigwedge_{i\in\mathcal{V}}\bigwedge_{j\in\mathcal{I}_i}\left(\hat{q}_k^{i,j}\geq 0  \right),\qquad k\in \mathbb{N},
\end{equation}
Furthermore, \textit{\textit{FRS}} tendency probabilities sum up to $1$ at every discrete time $k$. Therefore 
\begin{equation}\label{oneqh}
    \bigwedge_{j\in\mathcal{V}}\left(\sum_{i\in\mathcal{I}_j}\hat{q}_k^{i,j}=1  \right),\qquad k\in \mathbb{N}.
\end{equation}

\subsection{Optimization}\label{Optimization}
We obtain  $\hat{q}_k^{i,j}$ by solving the following optimization problem:
\begin{equation}
    \min {1\over 2} \sum_{i\in \mathcal{V}}\sum_{j\in \mathcal{I}_i}\hat{q}_k^{i,j}\hat{q}_k^{i,j}
\end{equation}
subject to constraint equations \eqref{upperconst}, \eqref{lowerconst},  \eqref{postiveqh},  and \eqref{oneqh}.

\begin{figure}[ht]
    \centering
    \includegraphics[width=\linewidth]{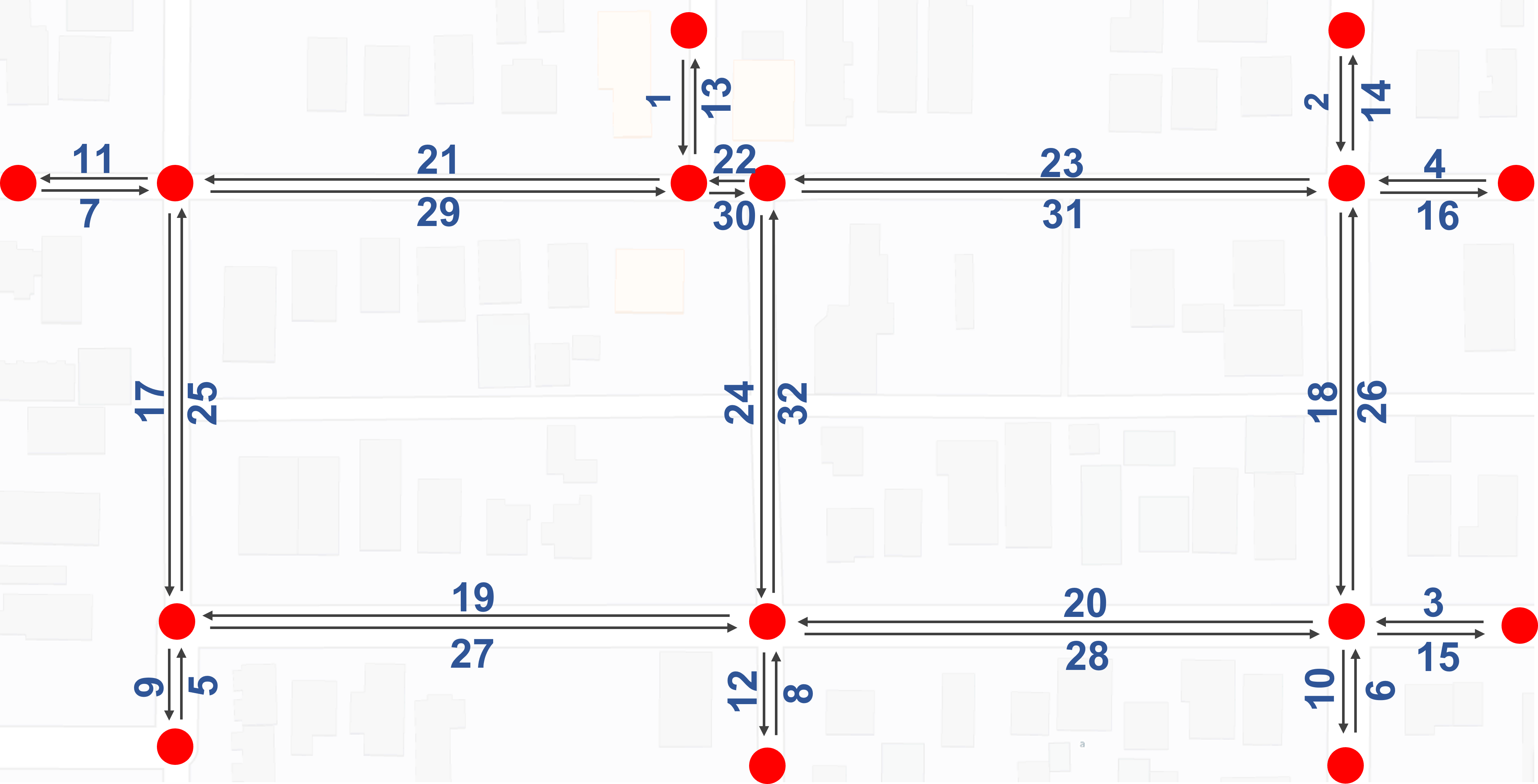}
    \vspace{-.3cm}
    \caption{Example NOIR: Street map of Phoenix.
    Numbers in blue denote the unidirectional roads $\mathcal{V}=\{1,\cdots,32\}$.}
    \label{realmap}
\end{figure}

\begin{table*}[t]
\caption{Road elements in the Phoenix city NOIR\cite{uppaluru2022continuous}}
\begin{adjustbox}{width=0.9\linewidth, center}
    \centering
\begin{tabular}{|c|c|c|c|c|c|}
  \hline
    ID & Name & Direction & ID & Name & Direction  \\
   \hline
   1 & N10th St.(E McKinley St.- E Pierce St.)& S & 2 &N11th St.(E McKinley St.- E Pierce St.)& S \\
    \hline
    3 &E Fillmore St.(N11th St.- N12th St.)& W & 4 &E Pierce St.(N11th St.- N12th St.)& W \\
    \hline
    5 & N9th St.(E Fillmore St.- E Taylor St.)& N  & 6 &N11th St.(E Fillmore St.- E Taylor St.)& N  \\
    \hline
    7 &E Pierce St.(N7th St.- N9th St.)& E & 8 &N10th St.(E Fillmore St.- E Taylor St.)& N \\
    \hline
    9 & N9th St.(E Fillmore St.- E Taylor St.)& S & 10 &N11th St.(E Fillmore St.- E Taylor St.)& S \\
    \hline
    11 &E Pierce St.(N7th St.- N9th St.)& W  & 12 &N10th St.(E Fillmore St.- E Taylor St.)& S \\
    \hline
    13 & N10th St.(E McKinley St.- E Pierce St.)& N & 14 &N11th St.(E McKinley St.- E Pierce St.)& N \\
    \hline
    15 &E Fillmore St.(N11th St.- N12th St.)& E & 16 &E Pierce St.(N11th St.- N12th St.)& E \\
    \hline
    17 & N9th St.( E Pierce St.- E Fillmore St.)& S & 18 & N11th St.( E Pierce St.- E Fillmore St.)& S \\
    \hline
    19 & E Fillmore St.(N9th St.- N10th St.)& W & 20 &E Fillmore St.(N10th St.- N11th St.)& W \\
    \hline
    21 & E Pierce St.(N9th St.- N10th St.)& W  &22 &E Pierce St.(N10th St.- N11th St.)& W \\
    \hline
    23 & E Pierce St.(N10th St.- N11th St.)& W  & 24 &N10th St.(E Pierce St.- E Fillmore St.)& S \\
    \hline
    25 & N9th St.( E Pierce St.- E Fillmore St.)& N & 26 &N11th St.( E Pierce St.- E Fillmore St.)& N \\
    \hline
    27 & N11th St.( E Pierce St.- E Fillmore St.)& S & 28 &N12th St.(E Pierce St.- E Fillmore St.)& N \\
    \hline
    29 & E Pierce St.(N9th St.- N10th St.)& E & 30 &E Pierce St.(N10th St.- N11th St.)& W \\
    \hline
    31 &E Pierce St.(N10th St.- N11th St.)& E & 32 &N10th St.(E Pierce St.- E Fillmore St.)& N \\
    \hline
\end{tabular}
\end{adjustbox}
\label{phoenix_road_name}
\end{table*}

\section{Simulation Results}\label{Simulation Results}
\begin{figure}[ht]
    \centering
    \includegraphics[width=\linewidth]{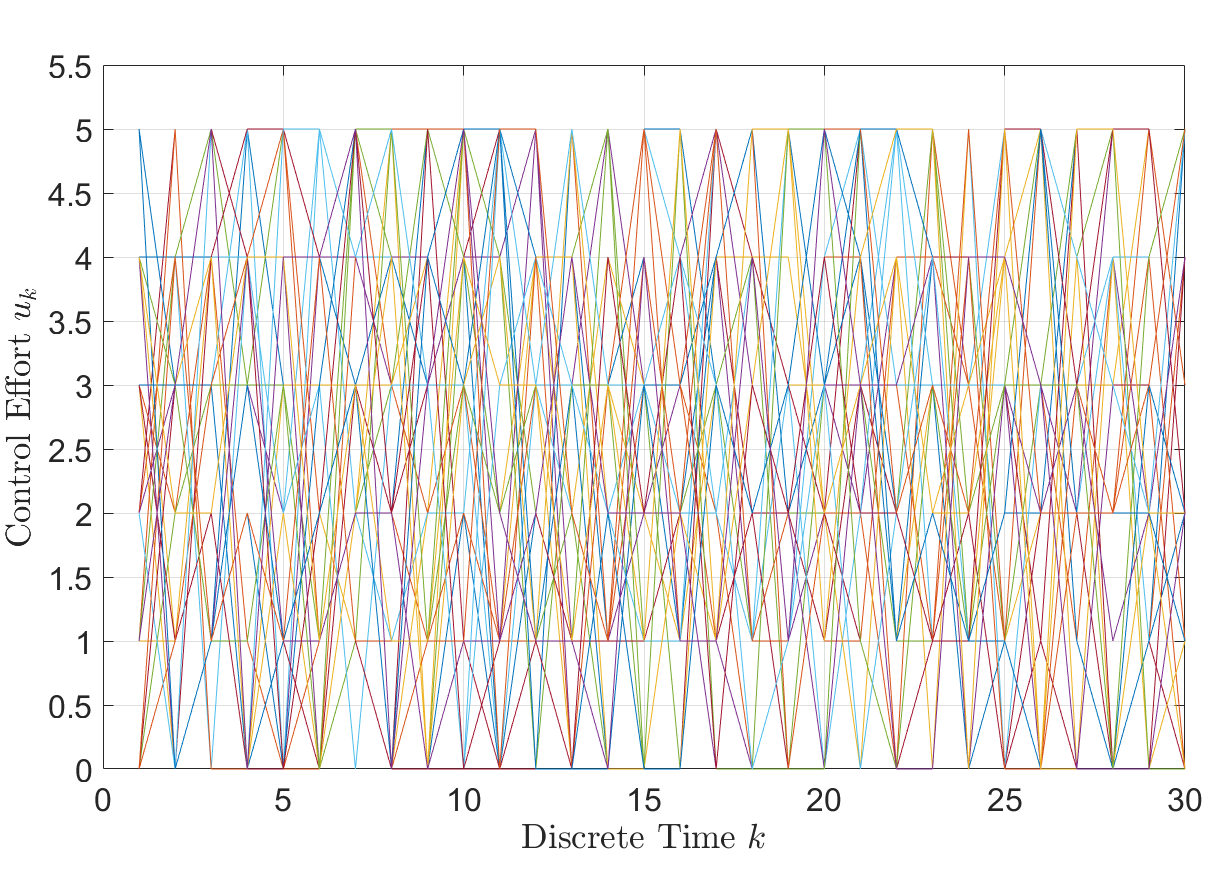}
    \vspace{-.3cm}
    \caption{Exogenous cars versus discrete time $k$ at every road $i\in \mathcal{V}$.}
    \label{exogen}
\end{figure}

\begin{figure}[ht]
    \centering
    \includegraphics[width=\linewidth]{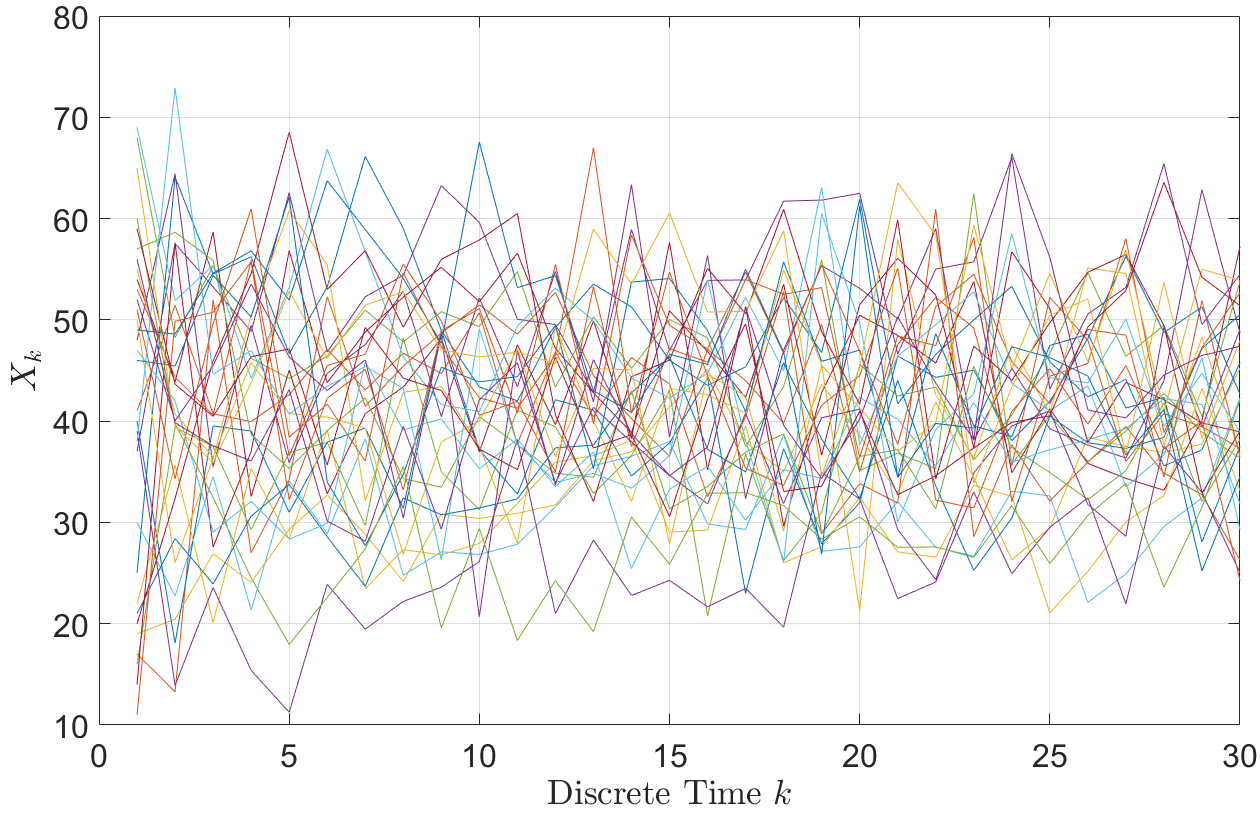}
    \vspace{-.3cm}
    \caption{Density of \textit{all} cars versus discrete time $k$ at every road $i\in \mathcal{V}$.}
    \label{u}
\end{figure}

\begin{figure}[ht]
    \centering
    \includegraphics[width=\linewidth]{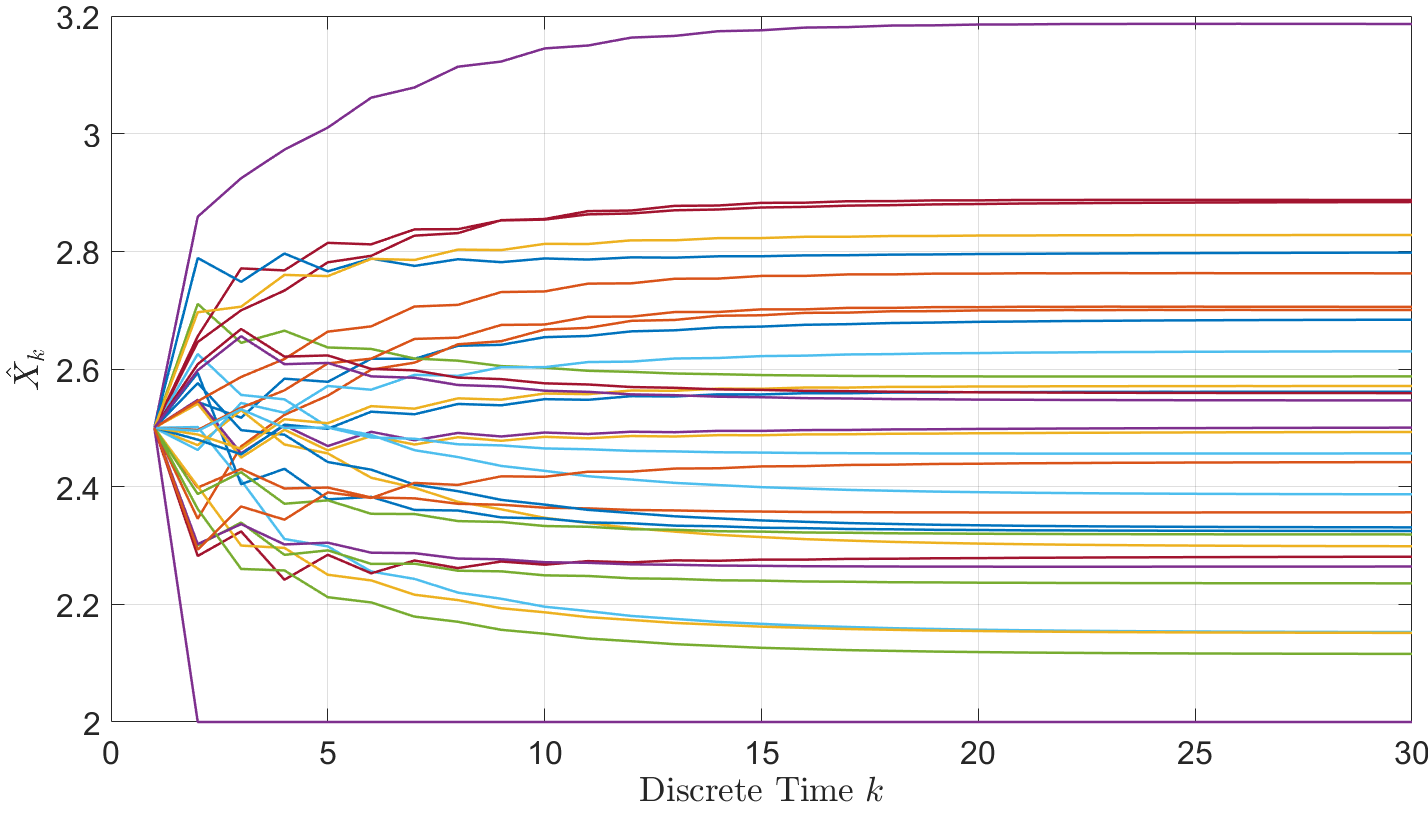}
    \vspace{-.6cm}
    \caption{Density of \textit{\textit{FRS}} cars versus discrete time $k$ at every road $i\in \mathcal{V}$.}
    \label{FRS}
\end{figure}

We simulate equitable distribution of FRS cars in a selected area in Downtown Phoenix with the map and NOIR shown in Fig. \ref{realmap}. The NOIR consists of $32$ unidirectional roads listed in Table \ref{phoenix_road_name} with the identification numbers defined by set $\mathcal{V}=\left\{1,\cdots,32\right\}$.  To ensure an equitable distribution of the \textit{FRS} cars, we require  that at least two \textit{FRS} cars are  available at every road $i\in \mathcal{V}$, therefore,
\[
\hat{\mathbf{x}}_{min}=2\mathbf{1}_{32\times 1}.
\]
We assume that exogenous cars enter and leave only through the inlet and outlet boundary nodes, respectively. Therefore, 
\begin{equation}
    d_k^i=\begin{cases}
        u_k^i\geq 0&i\in \left\{1,\cdots,8\right\}\subset \mathcal{V}\\
        -u_k^i\leq 0&i\in \left\{9,\cdots,16\right\}\subset \mathcal{V}\\
        0&\mathrm{otherwise}
    \end{cases}
    .
\end{equation}
Then, the traffic dynamics \eqref{allcardynamics} is converted to
\begin{equation}
    \mathbf{x}_{k+1}=\mathbf{A}_k\mathbf{x}_k+\mathbf{B}\mathbf{u}_k,
\end{equation}
where $\mathbf{u}_k=\begin{bmatrix}
    u_k^1&\cdots&u_k^{16}
\end{bmatrix}\mathbb{R}^{N_{out}\times 1}$ is the boundary control vector, and $\mathbf{B}=\left[b_{ij}\right]\in \mathbb{R}^{N\times N_{out}}$ is obtained by
\begin{equation}
    b_{ij}=\begin{cases}
        1&\left(i=j\right)\wedge \left(i\in \mathcal{V}_{in}\right)\\
        -1&\left(i=j\right)\wedge \left(i\in \mathcal{V}_{out}\right)\\
        0&\mathrm{otherwise}\\
    \end{cases}
    .
\end{equation}
For simulation, we use the MATLAB command `randsample.m'  to randomly assign $u_k^1$ through $u_k^{16}$ an integer from $0$ to $5$ at every discrete time $k$. Figure \ref{exogen} plots the boundary inputs $u_k^1$ through $u_k^{16}$ versus discrete time $k$. We also used the `randsample.m' command to generate a time-varying tendency probability matrix $\mathbf{Q}_k$ with time-invariant structure defined by graph $\mathcal{G}\left(\mathcal{V},\mathcal{E}\right)$. Additionally, the `randsample.m' command was used to generate a time-invariant, diagonal, and positive definite  outflow probability matrix $\mathbf{P}_k=\mathbf{P}\in \mathbb{R}^{32\times 32}$.
Figure \eqref{u} plots the density of \textit{all} cars versus discrete time $k$ for every road $i\in \mathcal{V}$. By using the approach proposed in Section \ref{Equitable Traffic Management} we optimized  $\hat{q}_k^{i,j}$ at every discrete time $k$, for every $i\in \mathcal{V}$ and $j\in \mathcal{I}_i$. The total number of \textit{FRS} cars, denoted by $\hat{x}_k^i$, is plotted versus discrete time $k$ in Fig. \ref{FRS} for every road $i\in \mathcal{V}$. It is seen the there exists at least $2$ \textit{FRS} cars in every road $i\in \mathcal{V}$ at all times, therefore, the inequality constraint \eqref{inequality} is satisfied.

\section{Conclusion}\label{Conclusion}
This paper developed a dissensus-based multi-agent coordination approach for evolution of networks under  a weighted network (Laplacian) matrix with columns summing up to $1$. We showed that the proposed dissensus dynamics can effectively apply CTM to model traffic evolution and integrate ride-sharing car usability. We also developed a constrained optimization problem to obtain an equitable distribution of ride-sharing cars in a congested traffic. We demonstrated  that equitable ride-sharing car distribution can be achieved by optimizing the tendency probability of FRS cars via solving a quadratic programming problem.

\IEEEpeerreviewmaketitle


\bibliographystyle{IEEEtran}
\bibliography{reference}

\end{document}